\documentclass{ws-procs9x6}

\usepackage{soul}

\def\etal {{\it et al.}}

\begin{document}

\title{EVIDENCE FOR SOLAR INFLUENCES ON NUCLEAR DECAY RATES}

\author{E.\ FISCHBACH$^{\dagger}$, J.H.\ JENKINS$^\ddagger$, J.B.\ BUNCHER$^\dagger$ and J.T.\ GRUENWALD$^\dagger$}
\address{$^\dagger$Physics Department, Purdue University,\\
$^\ddagger$School of Nuclear Engineering, Purdue University\\
West Lafayette, IN 47907, USA}

\author{P.A.\ STURROCK$^*$}
\address{Center for Space Science and Astrophysics, Stanford University,\\
Stanford, CA 94305, USA\\
$^*$E-mail: sturrock@stanford.edu}

\author{D.\ JAVORSEK II}
\address{416th Flight Test Squadron, 412th Test Wing, Edwards AFB,\\
Edwards AFB, CA 93524, USA}

\begin{abstract}
Recent reports of periodic fluctuations in nuclear decay data of certain isotopes have led to the suggestion that
nuclear decay rates are being influenced by the Sun, perhaps via neutrinos. Here we present evidence for the
existence of an additional periodicity that appears to be related to the Rieger periodicity well known in solar physics.
\end{abstract}

\bodymatter

\section{Introduction}\label{aba:Intro}

Our collaboration has recently produced evidence of small but significant
temporal changes 
in the decay rates of certain isotopes as a result of a mechanism presently unknown, but which
appears to be solar related.\cite{JenkinsFlare,JenkinsCorrel,Fischbach,JenkinsAnal,SturrockBNL,Javorsek} The
data which form the basis for this suggestion came from several sources. One of these comprised measurements of the decay rate of $^{54}$Mn, acquired at Purdue University in 2006, for which a decrease
in the measured count rate was coincident with the solar flare of 2006 December 13.\cite{JenkinsFlare,Fischbach} Further studies of data collected at Brookhaven National
Laboratory (BNL) measuring $^{32}$Si and $^{36}$Cl,\cite{Alburger,JenkinsCorrel,Fischbach,SturrockBNL,Javorsek} and
$^{226}$Ra data collected at 
the Physikalisch-Technische Bundesanstalt (PTB)\cite{Siegert,JenkinsCorrel,Fischbach,SturrockPTB} appear to
support this claim, in that the decay-rate data exhibit frequencies that appear to be related not only to the Sun-Earth distance, but also to solar rotation.
It should be emphasized that what is observed experimentally in each case is a deviation 
of the measured \textit{count rates} of the respective isotopes
from what would be expected by inserting the accepted half-lives into the familiar exponential decay law.

Of course, the fact that the measured \textit{count rates} exhibit an anomalous behavior does not 
necessarily imply that the intrinsic \textit{decay rates} are also anomalous, 
since systematic changes in the detector systems could be responsible
for the unexpected behavior. For example, the charge-collection efficiency of a gas detector system 
could be influenced by temperature, and hence be responding to small environmental (e.g. seasonal) 
changes in the ambient laboratory conditions. In what follows we present several arguments against 
a simplistic, systematic explanation of the BNL and PTB data fluctuations
in terms of environmental influences. When combined with similar arguments for the flare data 
in Refs.\ \refcite{JenkinsFlare} and \refcite{Fischbach}, we are led to suggest 
that nuclear decays may be intrinsically influenced by the Sun through some as-yet unexplained mechanism,
possibly involving neutrinos. We begin by summarizing the arguments against the proposition
that the observed effects in the decay rate 
measurements are due simply to environmental effects:
\begin{enumerate}
\item The apparent association between the solar flare of 2006 December 13 and a decrease in the 
$^{54}$Mn counting rate occurred over too short a time ($\sim43$ min) to be attributable
to any known seasonal environmental effect.\cite{JenkinsAnal}
\item In both the BNL experiment, which studied $^{32}$Si and $^{36}$Cl in the same detector,\cite{Alburger} 
and the CNRC (Children's Nutrition Research Center) experiment, which utilized $^{56}$Mn and $^{137}$Cs in 
the same detector,\cite{Ellis} the observed anomalies were different within each pair of isotopes. 
In the BNL experiment,
for example, ten 30-minute runs on $^{32}$Si were alternated with ten 30-minute runs on $^{36}$Cl to produce
a single data point for each of these nuclides on a given day. If the apparatus itself were solely responsible
for the observed annual fluctuations, then we would expect the fluctuations in the $^{32}$Si and $^{36}$Cl 
data to be the same, which they are not.\cite{JenkinsAnal,SturrockBNL}
\item In Ref.\ \refcite{JenkinsAnal}, a detailed analysis is presented of the effects of temperature, 
air pressure, and relative humidity fluctuations on the operation of the detectors used in the BNL and PTB
experiments. It is shown that the annual variations in these environmental factors were too small to
account for the observed annual fluctuations in the decay data.
\end{enumerate}
The preceding observations are not compatible with the observed effects being the 
result of systematic influences, and instead point to possible changes in
the intrinsic rate of the decay process. An even more compelling indication of an external influence, perhaps
of solar origin, arises from the discovery of additional
periodicities in the BNL and PTB data, which correspond to known solar periodicities,\cite{SturrockBNL,Javorsek,SturrockPTB}
but which are not seen in any environmental data. In Refs.\ \refcite{SturrockBNL} and \refcite{SturrockPTB},
it was shown that both the BNL and PTB data exhibited frequencies in the range 10-15 yr$^{-1}$, which are compatible with rotation frequencies appropriate for solar internal rotation. In what follows, we present evidence for another periodicity in both the BNL and PTB data, which appears to be related to the solar ``Rieger periodicity''.\cite{Rieger} This observation strengthens the case that the Sun
could be affecting terrestrial nuclear decays.

\section{Evidence for a Rieger-type Periodicity}
Apart from periodicities due to the solar cycle and to solar rotation, there is one more well known 
periodicity in solar data. This is the "Rieger" periodicity discovered in 1984 by Rieger and his 
colleagues in gamma-ray-flare data.\cite{Rieger} It has a period of about 154 days, corresponding to a 
frequency of 2.37 yr$^{-1}$. We have proposed that this may be interpreted as an r-mode frequency 
with spherical harmonic indices $l = 3, m = 1$.\cite{Sturrock99} The basic formula for these frequencies, as 
measured in a rotating fluid (the Sun), is
\begin{equation}
\nu(l,m)=\dfrac{2m\nu_R}{l(l+1)}
\label{eq1}
\end{equation}
\noindent where $\nu_R$ is the sidereal rotation frequency. This leads to the 
estimate $\nu_R=14.22~\textrm{yr}^{-1}$, which suggests that the oscillations 
are located in the transition region between the radiative zone and the convection zone (the tachocline).\cite{Schou}

We may now ask whether a similar oscillation occurs in (or perhaps near) the solar core, and whether 
this oscillation is manifested in decay data. We have found a periodicity at 11.93 yr$^{-1}$ in BNL 
data, one at 12.11 yr$^{-1}$ in PTB data, and one at 11.85 yr$^{-1}$ in a combined analysis of Homestake 
and GALLEX neutrino data and ACRIM irradiance data.\cite{Sturrock08,Sturrock09} This leads us to adopt a search band 
of 11 to 12.5 yr$^{-1}$ for a synodic rotation frequency, which converts to a sidereal rotation 
frequency of 12 to 13.5 yr$^{-1}$. These estimates are lower than the estimated rotation frequency 
of the radiative zone (13.9 yr$^{-1}$), indicative of a slowly rotating core.

\begin{figure}
\begin{center}
\epsfig{file=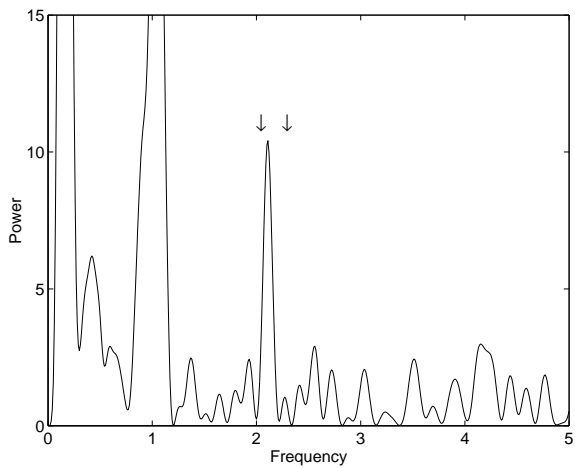,width=3.0in}
\end{center}
\caption{Section of the power spectrum of BNL data.}
\label{fig1}
\end{figure}

\begin{figure}
\begin{center}
\epsfig{file=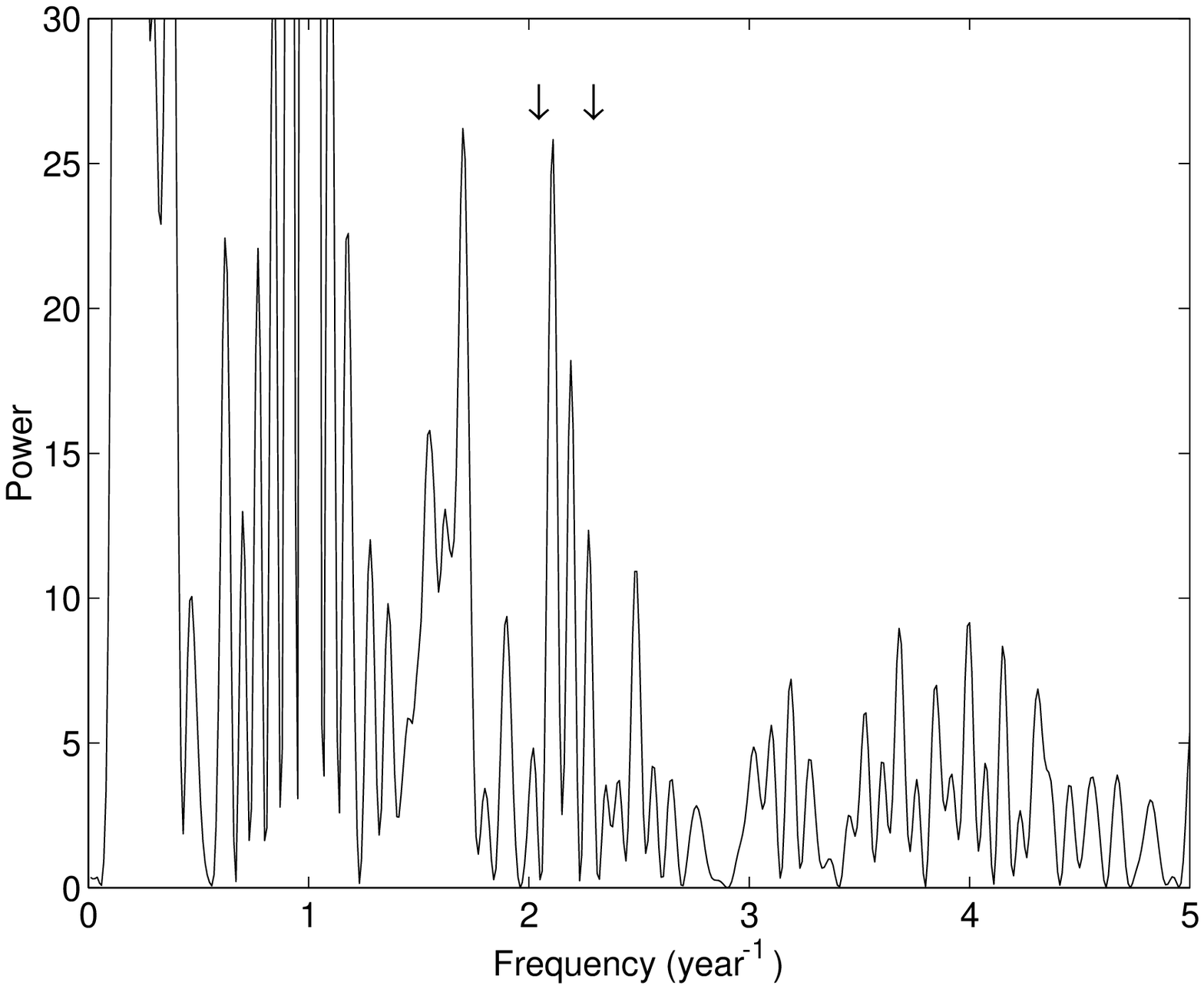,width=3.0in}
\end{center}
\caption{Section of the power spectrum of PTB data.}
\label{fig2}
\end{figure}

\begin{figure}
\begin{center}
\epsfig{file=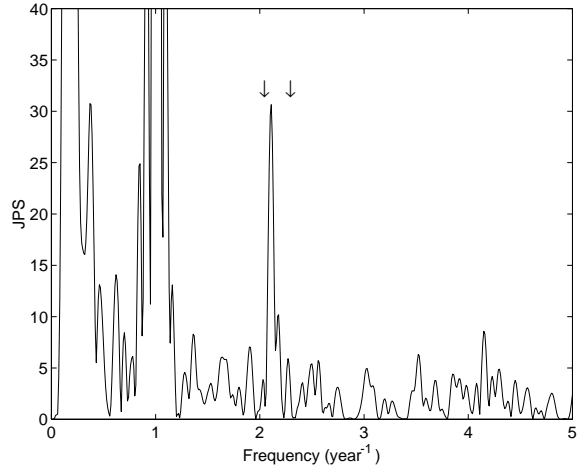,width=3.0in}
\end{center}
\caption{The joint power statistic formed by combining the BNL and PTB power spectra.}
\label{fig3}
\end{figure}

We therefore examine BNL and PTB data for evidence of a "Rieger-like" oscillation with a frequency 
given by Eq.\ \ref{eq1} with $l = 3, m = 1$, and $\nu_R$ in the range 12 to 13.5 yr$^{-1}$, which leads to the 
search band 2.00 to 2.25 yr$^{-1}$. On examining the power spectra shown in Figs.\ \ref{fig1} and \ref{fig2}, we 
find a peak in the BNL power spectrum at 2.11 yr$^{-1}$ with power $S$ = 10.09, and one in the PTB power 
spectrum at precisely the same frequency with $S$ = 25.83. When we combine the two power spectra by 
forming the joint power statistic $J$\cite{Sturrock05} (Fig.\ \ref{fig3}), we obtain $J$ = 30.65 at that frequency.

\begin{figure}
\begin{center}
\epsfig{file=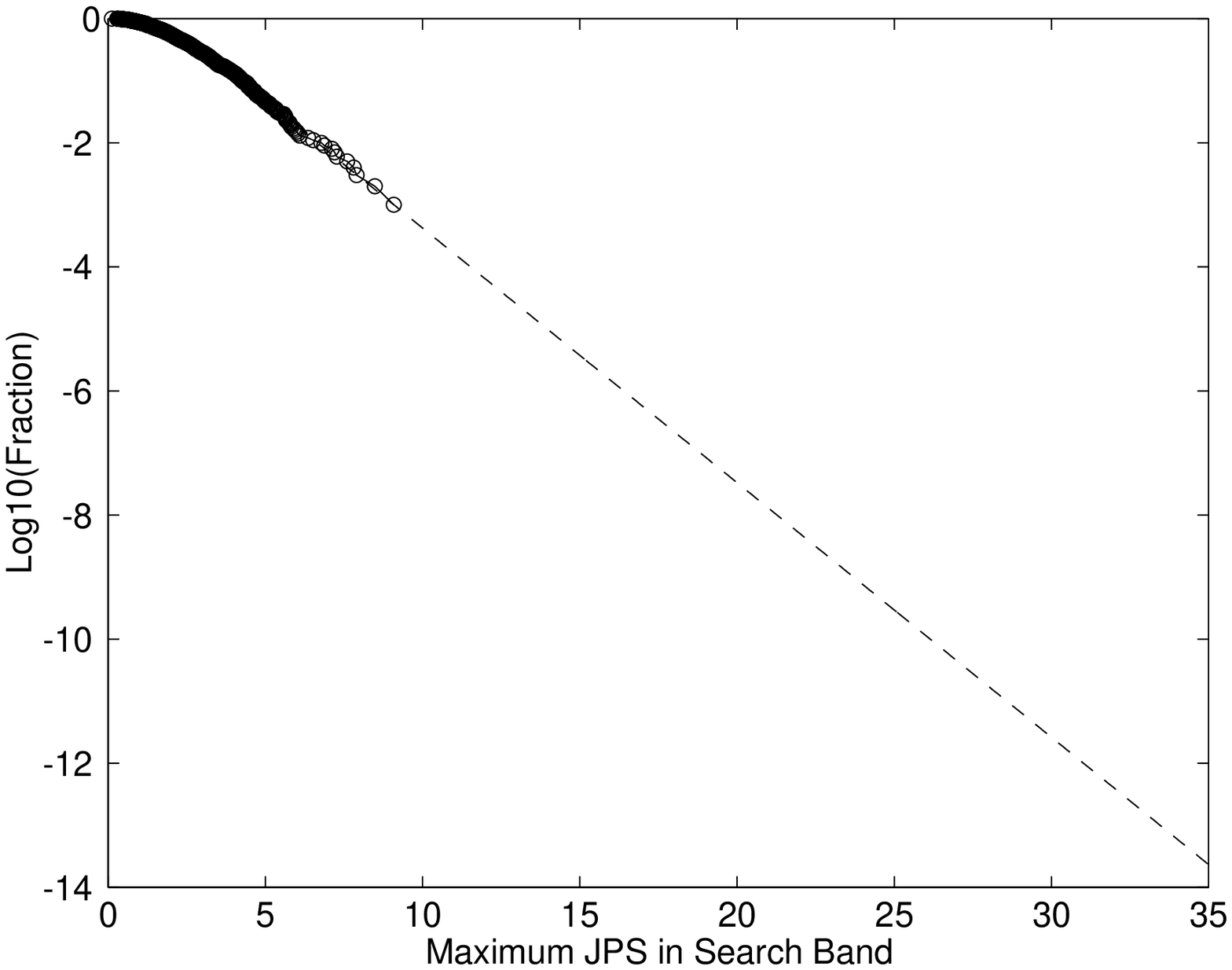,width=3.0in}
\end{center}
\caption{Logarithmic display of the results of the shuffle test applied to the joint power statistic. There is negligible probability of obtaining by chance a value as large as or larger than the actual value (30.65).}
\label{fig4}
\end{figure}

In order to assess the significance of this result, we have computed $J$ for 10,000 Monte Carlo 
simulations generated by the shuffle procedure,\cite{Bahcall} and for 10,000 simulations generated by the 
shake procedure,\cite{SturrockBNL} shuffling and shaking both datasets. The results from the shuffle test are 
shown in Fig.\ \ref{fig4}. The results of the shake test are virtually identical. These tests indicate 
that there is negligible probability of obtaining by chance a value of the JPS as large as or 
larger than the actual value (30.65).

This result appears to confirm our proposal that the Rieger periodicity is due to an r-mode 
oscillation, and to indicate that such an oscillation occurs in the solar core, influencing 
the solar neutrino flux and thereby influencing certain nuclear decay-rates.

\section*{Acknowledgments}
We are indebted to D. Alburger, G. Harbottle and H. Schrader for supplying us with their respective raw data.
The work of PAS was supported in part by the NSF through Grant AST-0097128,
and EF was supported in part by the U.S. DOE contract No. DE-AC02-76ER071428.
The views expressed in this paper are those of the
authors and do not reflect the official policy or position of the USAF, the US DOD, or the US Government.

\end{document}